\documentclass[titlepage,nofootinbib,reprint,twocolumn,showpacs,showkeys]{revtex4-2}

\usepackage{natbib}
\usepackage{amsmath}
\usepackage{amsfonts}
\usepackage{amssymb}
\usepackage{mathtools}
\usepackage[]{appendix}
\usepackage{setspace}
\usepackage{booktabs}
\usepackage{array}
\usepackage{multirow}
\usepackage{graphicx}
\usepackage{url}
\usepackage[USenglish]{babel}
\usepackage{graphics}

\newcolumntype{P}[1]{>{\centering\arraybackslash}p{#1}}

\begin{document}

%Title of paper
\title{Assessing the Impact of Social Network Structure\\ on the Diffusion of Coronavirus Disease (COVID-19): \\ A Generalized Spatial SEIRD Model}
 
%\title{Social Network Structure and Diffusion of Coronavirus Disease (COVID-19) in a Generalized Spatial SEIR Model}

\author{Giorgio Fagiolo}
%\email[]{}

%\homepage[]{Your web page}
%\thanks{}
%\altaffiliation{}

\affiliation{Istituto di Economia\\ Scuola Superiore Sant'Anna\\Piazza Martiri della Libert\`a 33, I-56127 Pisa (Italy)\\E-Mail: giorgio.fagiolo@santannapisa.it\\ORCiD: 0000-0001-5355-3352}

\date{October 2020}

\begin{abstract}

\begin{center}
\textbf{\\ Abstract \\}
\end{center}
In this paper, I study epidemic diffusion in a generalized spatial SEIRD model, where individuals are initially connected in a social or geographical network. As the virus spreads in the network, the structure of interactions between people may endogenously change over time, due to quarantining measures and/or spatial-distancing policies. I explore via simulations the dynamic properties of the co-evolutionary process dynamically linking disease diffusion and network properties. Results suggest that, in order to predict how epidemic phenomena evolve in networked populations, it is not enough to focus on the properties of initial interaction structures. Indeed, the co-evolution of network structures and compartment shares strongly shape the process of epidemic diffusion, especially in terms of its speed. Furthermore, I show that the timing and features of spatial-distancing policies may dramatically influence their effectiveness. 
\end{abstract}

%\pacs{aaa}
\keywords{Corona Virus Disease; COVID-19; Diffusion Models on Networks; Spatial SEIRD Models.}

\maketitle
\singlespacing
\section{Introduction\label{sec:introduction}}
In the last months, the still ongoing diffusion of the Coronavirus (COVID-19) pandemia has spurred a large body of scientific contributions, attempting to explore how compartmental models \cite{Grassly_Fraser_2008,Keeling_Rohani_2008,Brauer_2008,Kiss_Simon_2017} can reproduce and predict the spread of the epidemics in different countries and regions \cite{kousha2020covid19,Adam_2020}. 

Most of this work has been focusing on models in which the mixing process between people in different states or compartments does not depend on the social or geographical space where they are embedded in. However, some previous literature has shown that the (complex) structure of networks describing the way agents can meet, and possibly get infected, may affect the dynamics of the epidemic diffusion and its long-run properties  \cite{Cardy_1985,Camacho_etal_1996,Keeling2005Networks,Lloyd_Valeika_2005,House_2012,Jin_etal_2014,Rusu_2015,Pastos_2015,Pellis_etal_2015}. Furthermore, as the virus spreads in the network, the structure of interactions between people may change over time, due to quarantining measures and/or spatial-distancing policies, which may possibly introduce a co-evolutionary effect dynamically linking disease diffusion and network properties \cite{Small_Cavanagh_2020,Perez_etal_2020,Block_etal_2020}.       

Motivated by these observations, the paper introduces a generalized spatial SEIRD model that, besides the standard four compartments (susceptible, exposed, infected, recovered, dead), also considers an additional `quarantined' state, i.e. a SEIQRD model \cite{Peng_etal_2020}. I explore how the properties of the spread of the epidemics depend on: (i) the structure of the social/geographic network initially connecting the agents in the population, which matches infective and susceptible agents; (ii) the evolution of the share of quarantined and recovered agents (as well as social distancing policies), which dynamically destroy or re-establish social links. 

More specifically, I play with a finite population of agents (i.e. nodes) initially placed on four different families of interaction structures: (a) regular 2-dimensional lattices with Moore neighborhoods; (b) small-world lattice \cite{Watts_Strogatz_1998}; (c) Erd\"{o}s-Renyi random graphs \cite{Erdos_Renyi_1960}; (d) scale-free (preferential-attachment) networks \cite{Barabasi_Albert_1999}. I then investigate via Monte-Carlo simulations how the epidemic diffusion is affected by network structures, as their initial average degree increases (which in turn makes their topological properties change) and as the coupled  dynamics of quarantined and recovered people deletes and restores social interaction links. Finally, I examine how alternative spatial-distancing policies, which are taking again center stage in the political and social debate as the second wave of COVID-19 rolls across Europe and elsewhere, interacts with the coevolutionary process of disease diffusion and network updating.    
                 
\section{Methods\label{sec:mats_and_meths}}

\subsection{A Simple Model without Spatial Distancing Policies\label{subsec:the_model}}
I begin describing a simple model where no spatial distancing policies are enforced. Consider a population $P$ of $N$ agents living in a city, which is initially isolated from other cities. Time is discrete and, for the only sake of convenience, I will use the terms `time periods' or `days' as synonyms. Agents physically interact according to a simple, undirected, binary graph without self-loops, which at time $t=0$ is defined as $G_0=(P,L_0)$ , where $P=\{1,\dots,N\}$ and $L_0$ is the initial edge list, defined as the set of pairs $(i,j)$ such that $i\neq j$, $i\in P$, $j \in P$, and $(i,j)\in L_0$ if and only if there exists an edge between $i$ and $j$ at $t=0$. The graph $G_0$ ---which, as we will see below, is going to evolve through time as the epidemics spreads--- can be considered as describing social or geographical links through which people normally meet friends or neighbors.

At time $t=0$ all nodes are in the state $S$ (susceptible), but a randomly-chosen share $\theta$ of them becomes exposed (i.e., $\lfloor \theta N \rfloor$ agents become in state $E$, due to a random inflow of infective agents from other cities). People in state $E$ enter in an incubation period without symptoms and are not infectious. 
At any $t>0$, I assume that each agent $i\in P$ meets all its neighbors, i.e. all $j \in V_{it}$, where $V_{it}=\{j \in P: (i,j)\in L_t\}$ and $L_t$ is the current edge list. In each time period, transitions between compartments (i.e., states) occur through a parallel updating mechanism according to the following rules:

\begin{enumerate}
	\item[(a)] An agent in state $E$ becomes in state $I$ (infective) after an incubation of $\lfloor D \rfloor$ time periods, where $D$ is an i.i.d random variable with probability distribution $p(D)$.   
	\item[(b)] An agent in state $E$ becomes infected with probability $\pi=1-(1-\alpha)^k$ if h/she meets $k$ infective agents in its neighborhood, where $\alpha$ is a parameter tuning the likelihood of becoming infected in a single direct meeting and $0 \leq k \leq |V_{it}|$.\footnote{In other words, $\pi$ is the probability of being infected by at least one infective neighbor in a random sequence of meetings.} 
	\item[(c)] An agent in state $I$ becomes quarantined (in state $Q$) with a daily quarantine rate (${DQR}_t$). Agents in state $Q$ cannot meet anyone, i.e. they instantaneously cut all their bilateral links with their neighbors.\footnote{Since I do not distinguish between mild and severe symptoms in the development of the illness, there is not any difference in the model between being quarantined at home or at the hospital.}	
	\item[(d)] An agent in state $Q$ dies (i.e., becomes in state $D$) with a daily death rate (${DRR}_t$), recovers (in state $R$) with a daily recovery rate (${DRR}_t$) or stays quarantined otherwise. Recovered agents are assumed to be immunized and re-establish connections that they used to have in $G_0$ (provided that neighbors are still alive and are not quarantined).        
\end{enumerate}    
A flow-chart description of model dynamics is provided in the Supplementary Material (SM), see Figure \ref{fig:model_flowchart}.   

\subsection{Initial Network Structures\label{subsec:model_alt_nets}}
The initial network $G_0$ is assumed to belong to one out of the following graph families:
\begin{enumerate}
	\item [(i)] Regular 2-dimensional boundary-less lattices endowed with the Chebyshev distance ($LA$ henceforth). This defines squared Moore neighborhoods of radius $r^{LA} \geq 1$ and degrees $k_i^{LA}=(2r^{LA}+1)^2-1$ for all $i$. 
	\item [(ii)] Small-worlds lattice \cite{Watts_Strogatz_1998} built starting from nodes placed on a ring, with rewiring probability $p^{SW}>0$ and expected average degree $\bar{k}^{SW}=2r^{SW}$, where $r^{SW} \geq 1$ is the interaction radius on the initial ring ($SW$ henceforth).
	\item [(iii)] Erd\"{o}s-Renyi random graphs \cite{Erdos_Renyi_1960}, with link probability $p^{ER}>0$ and expected average degree 
		$\bar{k}^{ER}=(N-1)p^{ER}$ ($ER$ henceforth).
	\item [(iv)] Scale-free networks with linear preferential-attachment \cite{Barabasi_Albert_1999} and entrance of $m^{SF} \geq 1$ new nodes, generating an expected average degree $\bar{k}=2m^{SF}+o(1/m^{SF})$ ($SF$ henceforth).   
\end{enumerate}

To summarize network topology, I focus, besides average degree, on three statistics that have been found to influence, in general, the spread of epidemics on graphs \cite{Lloyd_Valeika_2005}. These are: the standard deviation of node degree distribution ($s_k$), global clustering coefficient ($c$) and average path-length ($\ell$), computed ignoring infinite path-lengths between nodes of different components. Their expected values (with standard errors) are reported in SM, Table \ref{tab:theor_graph_properties}. To get a better feel, fixing $r^{LA}\in\{1,2,3,4\}$, and thus $\bar{k} \in \{8,24,48,80\}$, $s_k$, $c$ and $\ell$ approximately scale as $\bar{k}^{\beta}$, with $\beta>0$ for $s_k$ and $c$ and $\beta<0$ for $\ell$ in all networks.  

\subsection{Parameter Setup\label{subsec:pars_setup}}
All simulations refer to a population of $N=1024$ agents (chosen to build a square lattice with edge $L=32$) and a number of days $T$ sufficient to reach a steady state.

The epidemic parameters of the model are calibrated using data at the national level for Italy, made available by ``Dipartimento della Protezione Civile'', see \url{https://github.com/pcm-dpc/COVID-19/tree/master/dati-andamento-nazionale}, covering the period from February, 22nd onward. In the simulations, I assume for simplicity that ${DRR}_t=DRR$, ${DQR}_t=DQR$ and ${DDR}_t=DDR$ and, on the basis of empirical diffusion curves, I build three epidemic scenarios: (i) strong-impact scenario: $(DQR, DDR, DRR)=(0.20, 0.10, 0.10)$; (ii) mid-impact scenario: $(DQR, DDR, DRR)=(0.15, 0.07, 0.15)$; (ii) low-impact scenario: $(DQR, DDR, DRR)=(0.10, 0.04, 0.20)$ ---see SM, Section \ref{sec:calibration_epidemics} for more details. Since the theoretical infection probability in a single meeting cannot be directly observed, I play with values of $\alpha$ that are $(0.20,0.10,0.05)$, respectively in the three scenarios. The percentage $\theta$ of exposed agents in day 0 is set to $5\%$ throughout. The probability distribution $p(D)$ of incubation days is calibrated using results from \cite{Lauer_etal_2020_incubation}, who show that $D$ is log-normally distributed with parameters $(\mu,\sigma)=(1.621,0.418)$. 

As to initial network structures, I experiment with average degrees $\bar{k}\in \{8,24,48,80\}$. These values result from setting $r^{LA}\in\{1,2,3,4\}$. Therefore, it follows that $r^{LA}\in\{4,12,24,40\}$, $p^{ER} \in \{\bar{k}(N-1)^{-1}, \bar{k}=8,24,48,80 \}$ and $m^{SF} \in \{4,12,24,40\}$. See SM, Table \ref{tab:pars_setup}, for a summary of parameter setups.
 
\subsection{Monte Carlo Simulations and Statistics\label{subsec:montecarlo_setup}}
For each choice of model parameters, I independently run  $M=1000$ simulations. This Monte Carlo sample size is sufficient to get standard errors for across-simulation averages small enough to ensure that differences between averages are always statistically significant. 

In order to get insights about within-simulation model behavior, I keep track of several \textit{within-simulation} statistics, i.e. computed in each day of the epidemic diffusion. These include: population shares in each compartment, death and cure rates, the share of agents who become infected through meetings, and the four network metrics $\bar{k}$, $s_k$, $c$, and $\ell$ ---which change across time as the result of the evolution population shares in each compartment. Another statistics of interest is the population-average of the number of neighbors that each $I$ agent has infected daily ($\bar{\rho}$ henceforth; see SM, Section \ref{sec:r0} for details), which can be employed as a rough estimate of the basic reproduction number ($R_0$) of the epidemics. Finally, I will also look at the spatial correlation coefficients of compartments (SCCC), calculating, for each state $\{S,E,I,Q,R,D\}$ the fraction of all existing edges in the network whose endpoints end up being in the same state. 

To summarize the aggregate behavior of the model (i.e. across runs), the following set of additional statistics are computed: (i) peak-time of infections (PTI), defined as the first day in which the share of infected people reach its overall maximum; (ii) the shares of agents in states $\{S,I,R,D\}$ at the end of simulation (EoS) and at PTI; (iii) the sum over all compartments of SCCC at PTI; (iv) the EoS share of agents who become infected through meetings; (v) the values of network metrics $\bar{k}$, $s_k$, $c$ and $\ell$ at PTI. Furthermore, I provide an estimate of the first day after which $\bar{\rho}$ goes below one (cf. SM, Section \ref{sec:r0}).  

Monte Carlo averages of all the above summarizing statistics will then be compared across initial networks families, initial average degrees, and epidemiological setups. 

%Notice that, in addition to the trivial steady-states of the diffusion process (i.e., all $S$ without influx, all $R$ and all $D$) there may be many other stable configurations (e.g. with $S$ agents), depending on the evolution of the disease over the changing topological structure of the network. 
   
\section{Results\label{sec:results}}

\subsection{Anatomy of Within-Simulation Dynamics in a Benchmark Setup\label{subsec:anatomy_sims}}

\begin{figure*}[ht!]
	\centering
	\includegraphics[width=12cm,height=12cm]{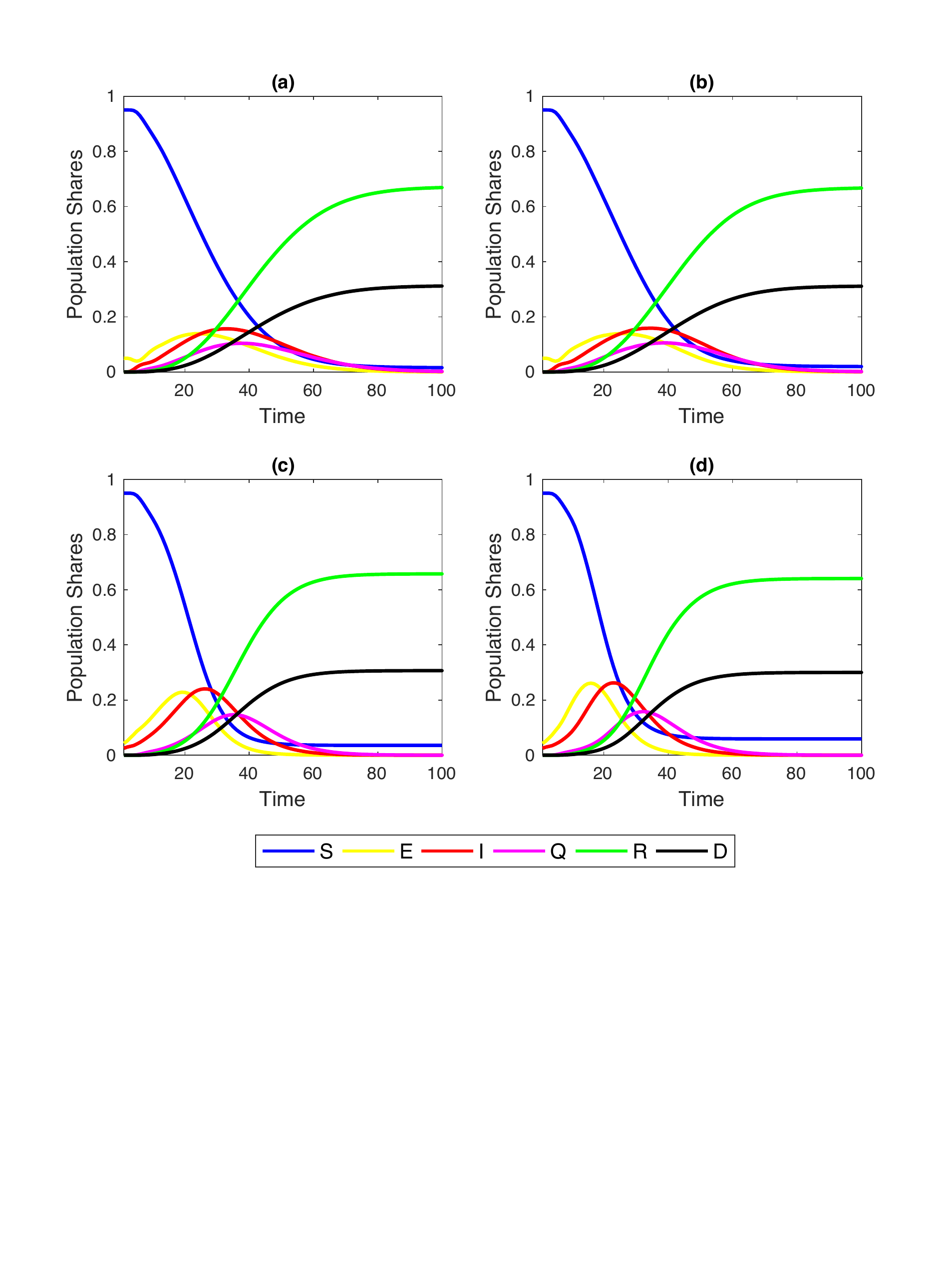}
	\caption{\label{fig:seiqrd_shares_time}Within-simulation evolution of agent shares in the six compartments over time. Initial $\bar{k}=8$. Mid-impact epidemic scenario. Averages across $M$=1000 Monte Carlo simulations. Panels: (a) regular 2-dimensional lattice with Moore neighborhoods; (b) small-world lattice; (c) Erd\"{o}s-Renyi random graph; (d) scale-free network. }
\end{figure*}

I begin studying the dynamic behavior of disease spreading across the four network families, focusing on the `Mid Impact' epidemic scenario with $\bar{k}=8$ (see Figure \ref{fig:seiqrd_shares_time}). Irrespective of the initial network structure, the population converges to a similar share of deaths, but in $ER$ and $SF$ networks a small percentage of $S$ people still remains (see below, Section \ref{subsec:alt_epidemics}).

In these two networks, epidemic diffusion reaches a higher peak of infections than in the case of $LA$ and $SW$, since more agents become exposed a little earlier. This is because in $ER$ and $SF$ networks the average number of infections per agent grows very quickly during the outbreak of the epidemic process, and then decreases earlier and more sharply than in $LA$ and $SW$ networks (SM, Figure \ref{fig:infections}). The evolution of SCCC shows, indeed, that the shares of edges linking two $E$ or two $I$ agents cross near to PTI and display a more abrupt inverse-U-shaped pattern over time, illustrating how the virus spreads across neighborhoods (SM, Figure \ref{fig:seiqrd_spat_corr}). 

As the epidemic process develops over time, the share of $Q$ agents first grows and then declines. This impacts on the network structure, because quarantined agents become isolated, constraining in turn the diffusion of the disease. Furthermore, the more the infection weakens, the more quarantined people recover and re-establish some of their initial connections. 

\vskip 1cm

\begin{figure*}[ht!]
	\centering
	\includegraphics[width=12cm,keepaspectratio]{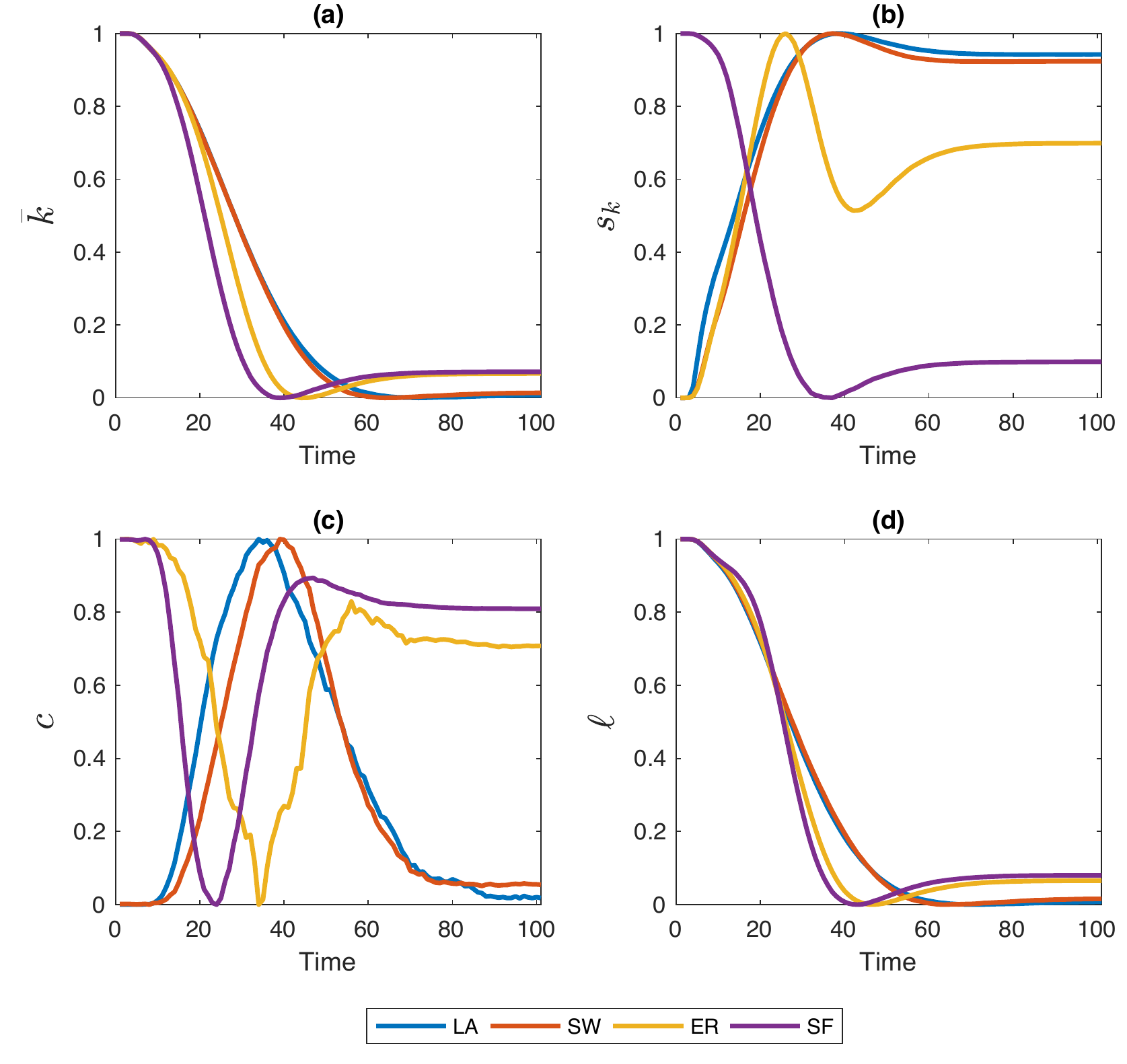}
	\caption{\label{fig:net_topology_time}Within-simulation evolution of network metrics, re-scaled to match the $[0,1]$ interval. Initial $\bar{k}=8$. Mid-impact epidemic scenario. Averages across $M$=1000 Monte Carlo simulations.}
\end{figure*}

To get a better feel about this co-evolutionary process, Figure \ref{fig:net_topology_time} shows how network metrics, normalized to match the $[0,1]$ interval, change during a simulation. Both $\bar{k}$ and $\ell$  decrease towards their minimum value across time in $LA$ and $SW$, with a pace slowing down as $R$ people spread in the population. The decline of $\ell$ is due to the growing number of small connected components and isolated nodes created by $Q$ and $D$ agents. In $LA$ and $SW$ networks, however, recovered agents that re-establish their connections are able to slightly boost average degree and reconnect isolated clusters. More marked differences across network structures emerge when looking at $s_k$ and $c$. In $LA$, $SW$ and, particularly, in $ER$ graphs $s_k$ first increases due to the injection of $Q$ agents, then as $R$ and $D$ gradually replace $Q$ patients, it oscillates until getting to a stable level. In $SF$ graphs, instead, $s_k$ follows the same time pattern of $\bar{k}$ and $\ell$, as initial heterogeneity is very high and cannot be further increased by the interplay between $Q$, $R$ and $D$ shares. Therefore, populations where the epidemics diffuse in $LA$, $SW$ and $ER$ networks end up having a higher final heterogeneity of degrees, while the opposite holds for $SF$ graphs. The final clustering level, instead, is almost completely recovered, but with opposite patterns. $LA$ and $SW$ networks first experience an increase in $c$ (albeit very moderate in magnitude) because the diffusion evolves less quickly. Instead, in $ER$ and $SF$ graphs, some triads are rapidly destroyed by $Q$ people and then $R$ people re-establish them when the epidemics softens. 

In the SM, Section \ref{sec:QD_NetStats}, I also show that, as the share of $Q$ agents first increases and then decreases, and that of $R$ agents keeps growing in time, the topological properties of the network changes in very heterogeneous ways, depending on the family to which it belongs. This is due to the dynamic removal and re-establishment of links ---which affects in non-trivial ways, in particular, the standard deviation of node degrees and global clustering coefficients--- and ultimately impacts on the properties of the diffusion process itself.

\subsection{The Impact of Initial Average Degree\label{subsec:impact_degree}}
I now investigate the behavior of the model when the initial average degree increases in the range $\{8,24,40,80\}$, keeping fixed the epidemic scenario to the `Mid Impact' one. If agents initially have, on average, more neighbors they can meet more infective people. Therefore, the probability to become $E$ increases for the population in each single day. However, a larger $\bar{k}$ does not imply the at the end of the simulation (EoS) there will be a larger fraction of deaths and/or recovered, as this is mainly affected by the epidemic parameters. What changes in the speed at which the contagion evolves and some of its dynamic properties.

\begin{figure*}[ht!]
	\centering
	\includegraphics[width=12cm,keepaspectratio]{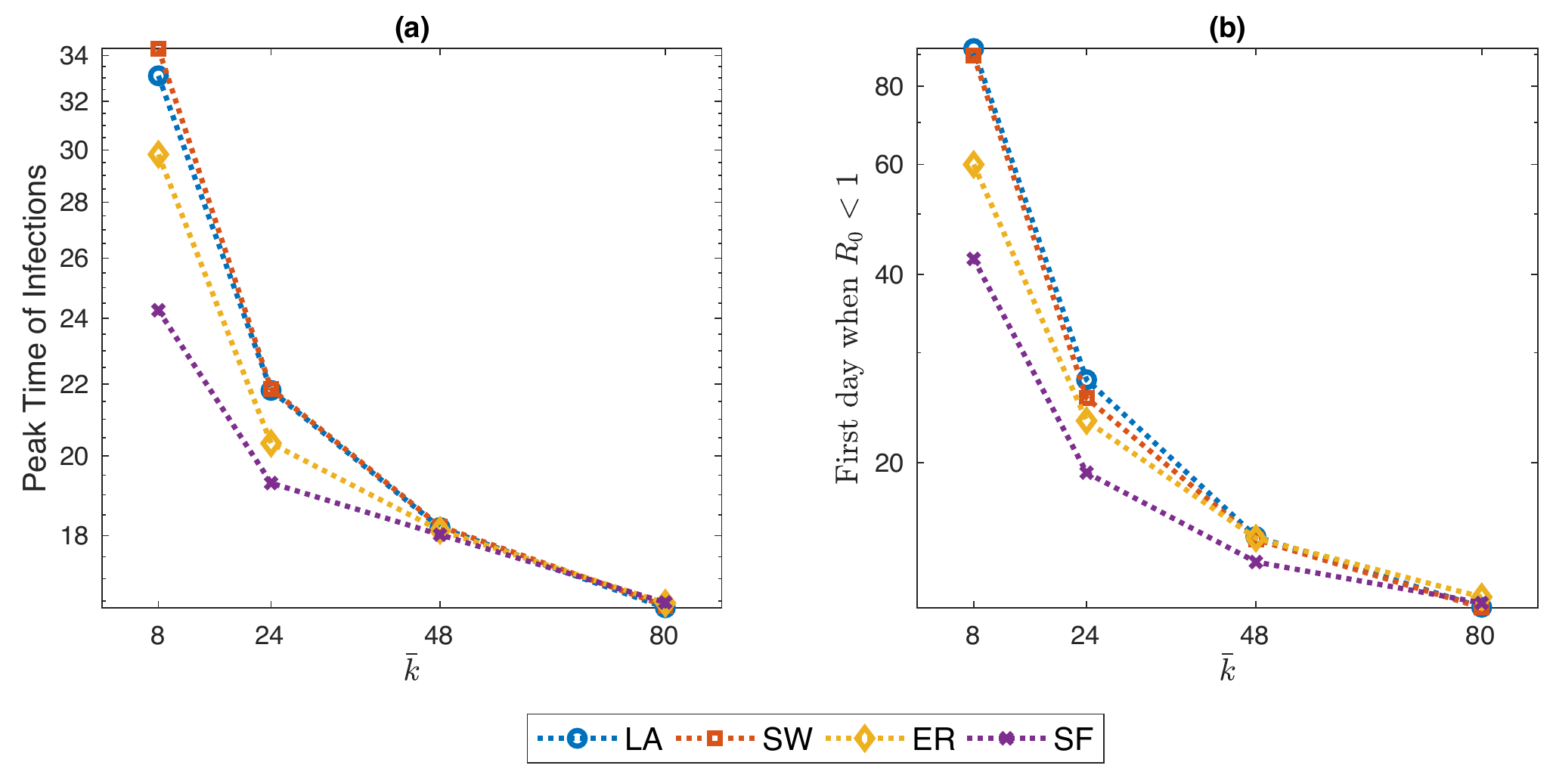}
	\caption{\label{fig:deg_speed}Panel (a): Peak-time of infections (PTI), defined as the first day in which the share of infected people reach its overall maximum, against initial average degree. Panel (b) estimate of the first day after which $\bar{\rho}$ goes below one (cf. SM, Section \ref{sec:r0}). Initial average degree in the range $\{8,24,40,80\}$. Mid-impact epidemic scenario. Averages across $M$=1000 Monte Carlo simulations. Y-axis in log scale.}
\end{figure*}

For example, as shown in Figure \ref{fig:deg_speed} both the peak-time of infections and the estimate of the first day after which $\bar{\rho}$ goes below one, quickly decrease with $\bar{k}$. Furthermore, as the initial average degree grows, the contagion evolves more quickly in ER and, especially, in SF networks.       
            
Furthermore, in all networks, the fraction of infected people at PTI immediately jumps up when $\bar{k}$ increases fromz 8 to 24, and then keeps growing with $\bar{k}$ but less quickly (cf. Figure \ref{fig:deg_shares} in the SM). This implies that, since the epidemic scenario is fixed, the share of agents that are quarantined in the first days of the contagion increases more than linearly. Therefore, at PTI, the shares of susceptible, recovered and dead agents actually decrease with initial average degree.

\subsection{Model Behavior in Alternative Epidemic Scenarios\label{subsec:alt_epidemics}}

\begin{figure*}[ht!]
	\centering
	\includegraphics[width=12cm,keepaspectratio]{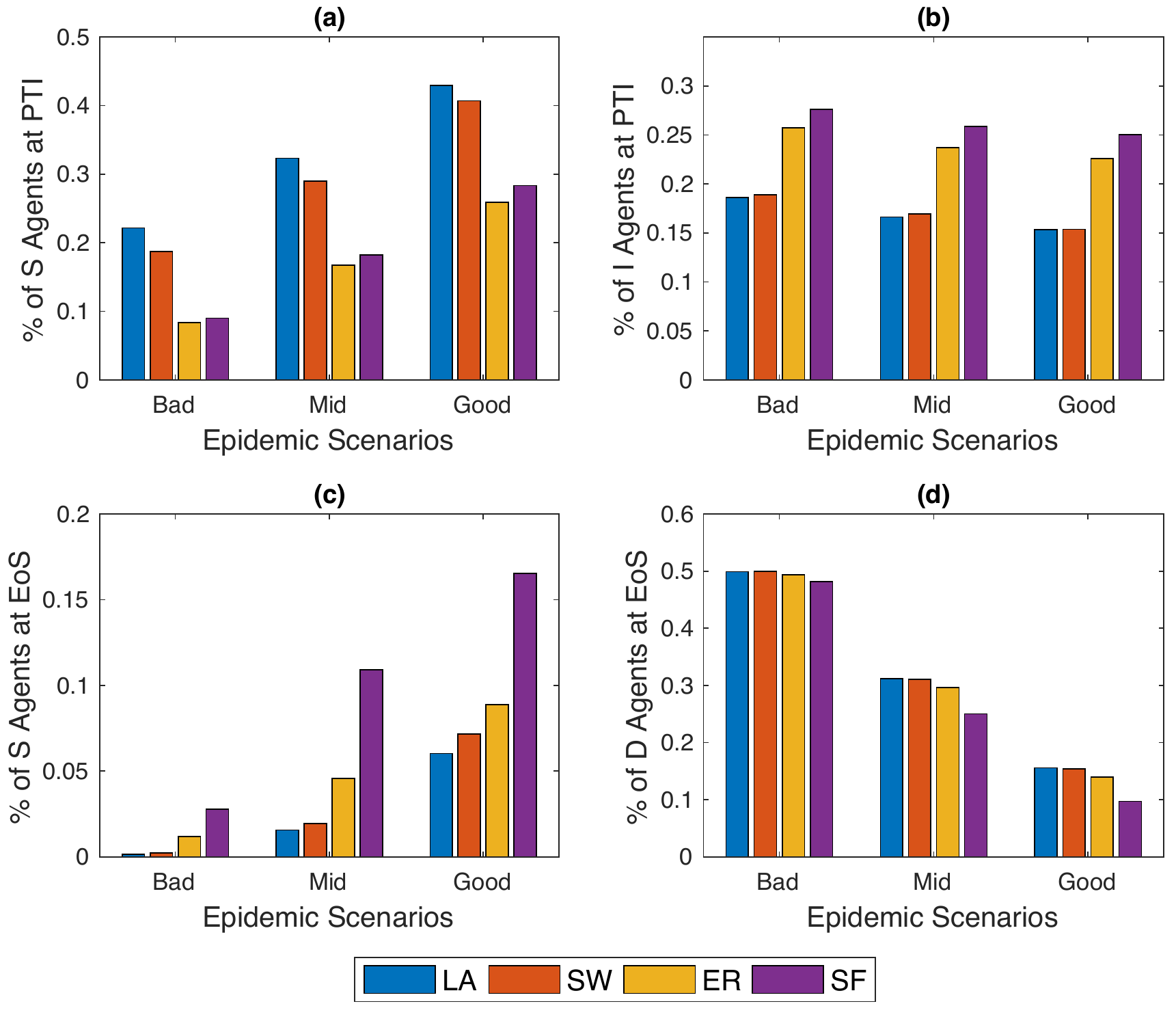}
	\caption{\label{fig:epid_scen_fig1}Comparing the behavior of the model across three epidemic scenarios: Bad vs. Mid vs. Good (see Section \ref{subsec:pars_setup}). Panels (a) and (c): \% of agents in compartment $S$ at peak-time of infections (PTI). Panel (b): \% of agents in compartment $I$ at the end of simulation (EoS). Panel (d): \% of agents in compartment $D$ at the end of simulation (EoS). Initial average degree: $\bar{k}=8$. Averages across $M$=1000 Monte Carlo simulations.}
\end{figure*}

Next, I explore what happens in the model when alternative epidemic scenarios are assumed (cf. Section \ref{subsec:pars_setup}). For the sake of comparison, I keep fixed $\bar{k}=8$ throughout. Simulation results show that, as expected, EoS shares of dead (respectively, recovered) agents decrease (respectively, increase) in all network setups as one moves from the bad to the good epidemic scenario (see Figure \ref{fig:epid_scen_fig1}). More interestingly, within the same scenario, the model behaves differently across network setups, and these differences are amplified as the contagion is less strong. Indeed, in ER and SF networks the epidemics diffuses quicker than in LA and SW graphs ---as documented in SM, Figure \ref{fig:epid_scen_fig2}, panels (a) and (c). Therefore, LA and SW display more $S$ (and less $I$) agents at PTI than ER and SW do --- see Figure \ref{fig:epid_scen_fig1}, panel (c)---  and a significantly smaller spatial correlation of compartments (panel (b) in SM, Figure \ref{fig:epid_scen_fig2}). At the end of the simulation (EoS), conversely, many more susceptible agents remain in ER and, especially, in SF networks. This is due to the higher heterogeneity of the degree distribution in such networks: the existence of many small-degree nodes at the beginning of the process prevents them to be infected, especially when the contagion becomes softer and their few neighbors are quickly quarantined. As a consequence, slightly smaller shares of deaths are observed in ER and, in particular, in SF networks at EoS.

\subsection{Spatial Distancing\label{subsec:spat_dist}}

 Spatial distancing (SD) is implemented in the model in a very stylized way. I assume that the city government only tracks the evolution of $Q$ agents and enforces SD when $x_t(Q)>q^\star$, where $x_t(Q)$ is the current share of agents in the $Q$ compartment and $q^\star \in (0,1)$. The SD policy aims at making more difficult face-to-face meetings between neighbors, and can be enforced with increasing strengths. Of course, its \textit{ex-post} effectiveness also depends on how strictly people follow the rules. Here, I do not separately model the \textit{ex-ante} plans of the government and the response of the agents. Therefore, more formally, I define $\theta \in (0,1)$ as the \textit{ex-post} effectiveness of SD policy and assume that, under SD, an agent meets each neighbor in any time period $t$ with probability $\psi=1-\theta$. This implies that, under SD, an agent in state $E$ now becomes infected with probability: 
 \begin{equation}
 	\pi^{SD}=1-(1-\psi\alpha)^k
 \end{equation} 
 where $k$ is the number of infective agents the agent meets in its neighborhood. I allow for two versions of SD: (i) \textit{permanent}, if SD is enforced from the first day when $x_t(Q)>q^\star$ onwards, i.e. during the period $\{\underline{t},\dots,T\}$, where $\underline{t}=\inf_t\{t:x_t(Q)>q^\star\}$; (ii) \textit{temporary}, if SD is enforced only whenever $x_t(Q)>q^\star$, and it is removed (i.e. $\theta$ is switched back to zero) if $x_t(Q)\leq q^\star-\epsilon$. Here, the $\epsilon$-term prevents the SD policy to be too sensitive to oscillations of $x_t(Q)$ around $q^\ast$, thus avoiding stop-and-go patterns. In the following simulations, I consider three SD setups: (a) strong: $(q^\star,\theta)=(0.02,0.7)$; (b) intermediate: $(q^\star,\theta)=(0.04,0.5)$; (a) mild: $(q^\star,\theta)=(0.06,0.3)$, whilst keeping fixed throughout $\bar{k}=8$ in the mid epidemic scenario and $\epsilon=0.05$.     

\begin{figure*}[ht!]
	\centering
	\includegraphics[width=12cm,keepaspectratio]{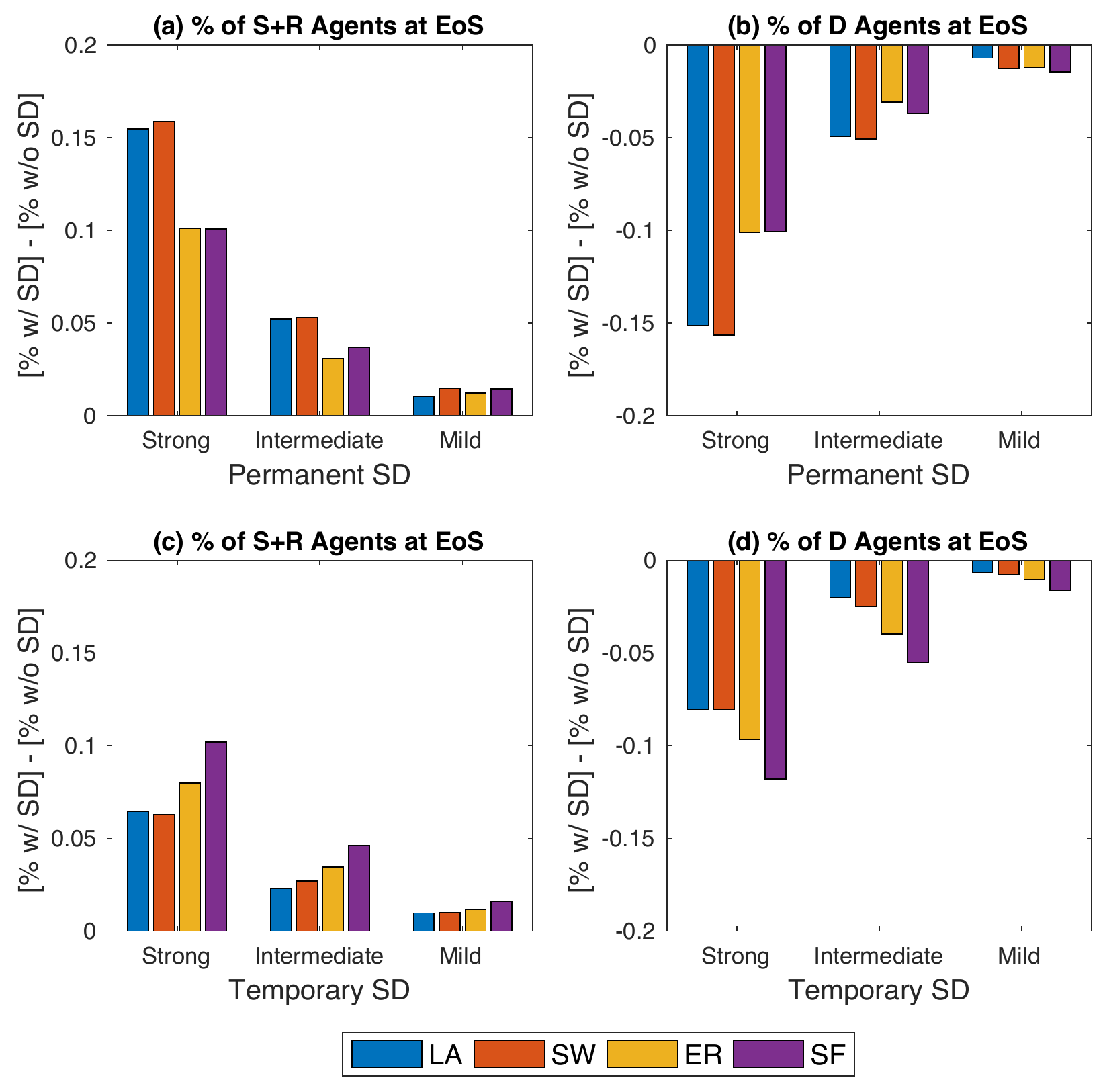}
	\caption{\label{fig:sd_fig1}Effects of spatial distancing (SD). Comparing the behavior of the model across three SD setups: Strong vs. Intermediate vs. Bad (see Section \ref{subsec:spat_dist}). Panels (a) and (b): Permanent SD policy. Panels (c) and (d): Temporary SD policy. Panels (a) and (c): Share of agents in states $S$ or $R$ at EoS with SD minus the same share without SD. Panels (b) and (d): Share of agents in state $D$ at EoS with SD minus same share without SD. Initial average degree: $\bar{k}=8$. Mid epidemic scenario. Averages across $M$=1000 Monte Carlo simulations.}
\end{figure*}
 
Figure \ref{fig:sd_fig1} plots EoS shares of agents under SD (either permanent or temporary) minus the correspondent share without SD. In each SD setup, I target the share of people ending up in either $S$ or $D$ compartments, and the share of deaths ($D$). Results show that, as expected, a permanent SD policy is better than a temporary one independently of network structure. However, especially when a strong setup is enforced in the permanent SD policy version, networked populations that benefit the most are those where agents are located on either lattices or small-worlds. Conversely, temporary SD policies are more effective in ER and, in particular, in SF networks, provided that they are implemented more rigorously.

This is due to how network structures evolve during a typical run, see Figure \ref{fig:net_topology_time}. Indeed, when a permanent SD policy is likely to be implemented, LA and SW exhibits larger average degrees and clustering than ER and SF. This prevents the infection to be transmitted more effectively during the peak. Instead, enforcing temporary SD policies allow an even smaller probability that low-degree agents remain susceptible, which is more likely to happen in ER and SF networks, due to their higher degree variability. When such a policy is switched off, ER and SF systems display higher (and more dispersed) average degrees and larger clustering than in the LA and SW cases, but the share of infected people is now smaller. Therefore, one observes less deaths. Disaggregating $S$ and $R$ shares shows also that, in the permanent SD case, the improvement in LA and SW is obtained via an almost similar increase of both compartments. On the contrary, when SD is temporary, much of the improvement is due to an increase in EoS susceptible agents only.

\section{Discussion\label{sec:discussion}}
In this paper, I studied a generalized spatial SEIRD model to explore the impact of alternative social-network structures on the diffusion of the COVID-19 disease. The introduction of quarantined agents generates a coevolving process between epidemic spreading and network structure, ultimately shaping steady-state outcomes and the speed of diffusion. 

In the simplest framework, without spatial distancing policies and a given benchmark choice of initial average degrees and epidemic parameters, the initial network structure does not affect the final shares of susceptible, dead and recovered people, but it strongly impact on the timing and the speed of diffusion. In ER and, in particular, in SF networks, more agents become exposed earlier and diffusion takes place quicker and more strongly than in the LA and SW cases. This is linked with how network structure coevolves across time with the shares of $Q$, $R$ and $D$ agents. Indeed, in ER and SF networks, average degree initially decreases less sharply than it does in LA and SW. Furthermore, degree variation and clustering is higher. Therefore, the probability of becoming exposed increases, as susceptible agents face larger and more clustered neighborhoods. Increasing initial average degree, while keeping fixed epidemic parameters, thus results in a faster speed of infection, especially in ER and SF networks, both in terms of smaller PTIs and average number of neighbors that each agent has infected daily. When instead different epidemic scenarios are assumed for a fixed initial degree, network structure impacts differently model behavior, and these differences are amplified as the strength of the contagion weakens. In particular, since the epidemics initially diffuses quicker in ER and SF networks, one typically observes more $S$ (and less $I$) agents at PTI in LA and SW graphs, and many more remaining $S$ agents at EoS in ER and SF networks (with slightly smaller shares of deaths).            

The effect of SD policies depends in the model on the strength with which they are enforced, as well as whether they are temporary of permanent. In particular, whereas permanent SD policies allow for better results than temporary ones irrespective of network structure, permanent (and strong) SD measures are more effective in LA and SW structures, whereas temporary (and strong) SD policies should be preferred if interactions occur through ER or SF graphs. This is again due to the interplay between network structure and compartment shares in the evolution of the epidemics. Indeed, switching on and off SD policies may hit the system when the topological properties of its network structure are very different, depending on the initial graph family describing social interactions.

More generally, results suggest that, in order to predict how epidemic phenomena evolve in networked populations, it is not enough to focus on the properties of initial interaction structures. In fact, if the epidemic diffusion requires quarantining people, and possibly enforcing SD policies, the coevolution of network structures and compartment shares strongly shape the way in which the virus spreads into the population, especially in terms of its speed. On the one hand, the average and standard deviation of degree distribution, as well as clustering, of initial networks are, together with epidemic parameters, important determinants of the subsequent diffusion patterns. On the other hand, the topology of social interaction structures evolves over time, due the rise and fall of $Q$, $R$ and $D$ agents, in different and non trivial ways across alternative network families, and this in turn impacts diffusion patterns. As a result, the timing and features of SD policies may dramatically influence their effectiveness.

The foregoing analysis can be extended and improved in several directions. To begin with, alternative parametrizations for the epidemic process, more in line with evidence from the ongoing second wave, could be tested. Furthermore, it would be interesting to assess the extent to which results are robust to increasing population size, additional network structures (e.g., core-periphery graphs), and different values for the share of agents that become initially exposed. In this last respect, one could also play with alternative assumptions as to the mechanism governing the way in which exposures initially occur, e.g. allowing for the emergence of spatially-clustered exposed agents, instead of just supposing that a randomly-chosen share of people get infected. Finally, one can perform a deeper analysis to better understand how the topology of network structures influences epidemic diffusion, for example asking whether centrality indicators such as k-coreness measures \cite{Bae_Kim_2014,Kitsak_etal_2010} can help in investigating the role of super spreaders \cite{superspreaders_1}.

%\clearpage 

%\bibliographystyle{aipnum4-2.bst}

%\bibliographystyle{unsrtnat}

%\bibliography{bib_spatial_seird}

%%%%%%%%%%%%%%%%%%%%%%%%%%%%%%%%%%%%%%%%%%%%%%%%%%%%%%%%
%%%% Supplementary material
%%%%%%%%%%%%%%%%%%%%%%%%%%%%%%%%%%%%%%%%%%%%%%%%%%%%%%%%

\clearpage
\onecolumngrid
\setcounter{page}{1}
\setcounter{figure}{0}
\setcounter{table}{0}

\begin{appendices}

\renewcommand{\thepage}{S\arabic{page}}  
\renewcommand{\thesection}{S\arabic{section}}   
\renewcommand{\thetable}{S\arabic{table}}   
\renewcommand{\thefigure}{S\arabic{figure}}

\begin{center}
	\huge{\textbf{Supplementary Material}}
\end{center}

\vskip 2cm

\section{Additional Figures and Tables\label{sec:add_figs_tabs}}

\vskip 2cm

\begin{figure}[hb]
	\includegraphics[width=16cm]{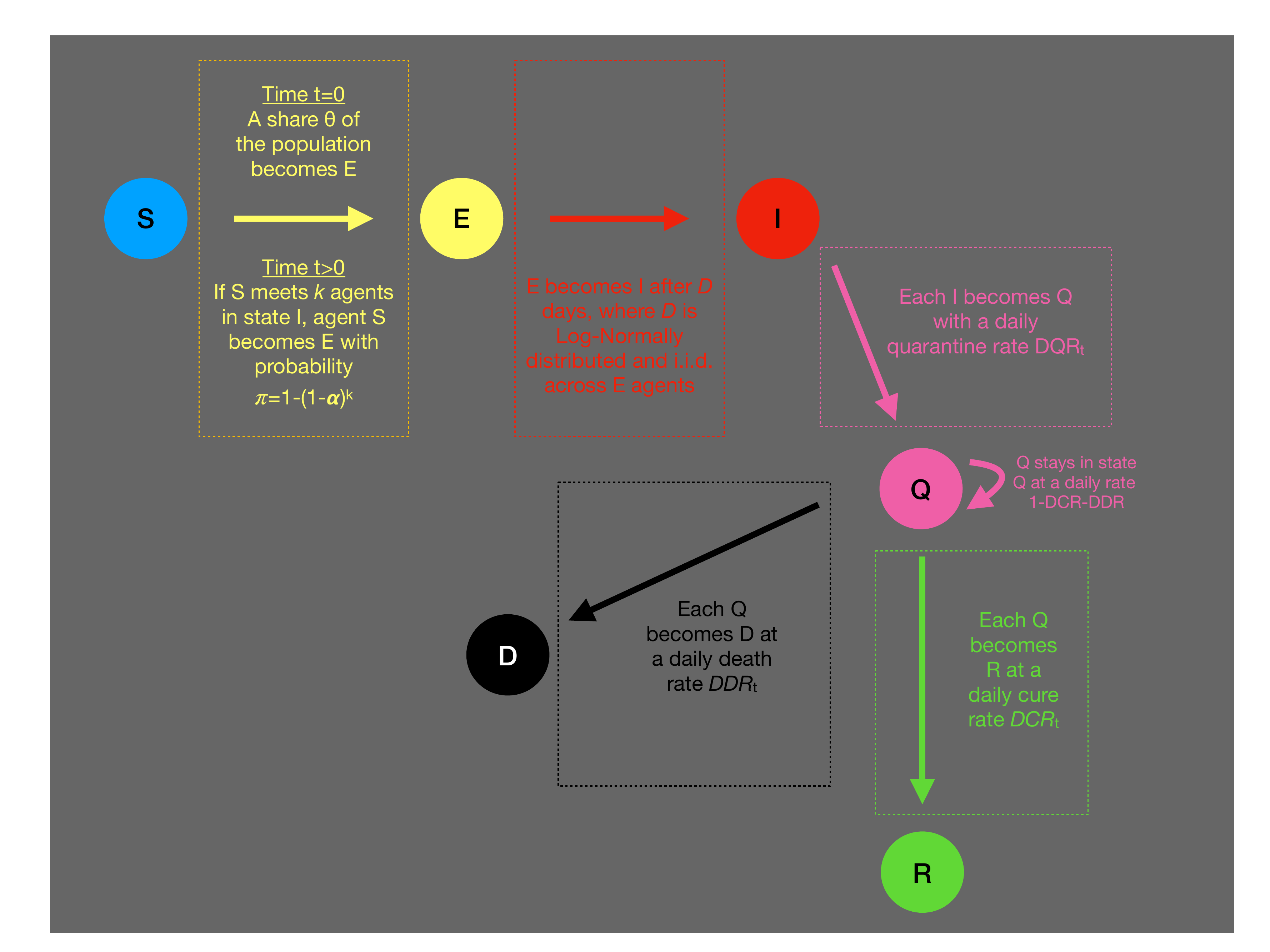}
	\caption{\label{fig:model_flowchart}Flow-chart of model dynamics in any given time period.}
\end{figure}

\clearpage \newpage

\setstretch{1.0}

\begin{table}[hb]
\centering
\resizebox{12cm}{!}{
	\begin{tabular}{P{3cm}|P{3cm}P{3cm}P{3cm}P{3cm}}
	\toprule
			&	\multicolumn{4}{c}{\textbf{Standard Deviation of Degree} ($s_k$)}			\\[-6pt]
\textbf{Avg Deg ($\bar{k}$)}&	Lattice	&	Small World	&	Erd\"{o}s-Renyi	&	Scale-Free	\\
\midrule
\multirow{2}{*}{8}	&	0.0000	&	0.8730	&	2.8180	&	8.8610	\\
	&	-	&	(0.00086)	&	(0.00207)	&	(0.01083)	\\[-6pt]
\multirow{2}{*}{24}	&	0.0000	&	1.5080	&	4.8330	&	21.2190	\\
	&	-	&	(0.00119)	&	(0.00349)	&	(0.01066)	\\[-6pt]
\multirow{2}{*}{48}	&	0.0000	&	2.1320	&	6.7570	&	36.6450	\\
	&	-	&	(0.00159)	&	(0.00489)	&	(0.0109)	\\[-6pt]
\multirow{2}{*}{80}	&	0.0000	&	2.7460	&	8.5820	&	53.9910	\\
	&	-	&	(0.002)	&	(0.00642)	&	(0.01076)	\\[-6pt]
	\midrule
\multirow{2}{*}{$\hat{\beta}$}	&	-	&	0.4978	&	0.4809	&	0.7766	\\
	&	-	&	$R^2$=1.0000 	&	$R^2$=0.9999	&	$R^2$=0.9999	\\[-6pt]
	\toprule
	&	\multicolumn{4}{c}{\textbf{Global Clustering Coefficient} ($c$)}		\\[-6pt]
\textbf{Avg Deg ($\bar{k}$)}	&	Lattice	&	Small World	&	Erd\"{o}s-Renyi	&	Scale-Free	\\
	\midrule
\multirow{2}{*}{8}	&	0.4290	&	0.4630	&	0.0080	&	0.0250	\\
	&	-	&	(0.00025)	&	(0.00003)	&	(0.00004)	\\[-6pt]
\multirow{2}{*}{24}	&	0.5220	&	0.5190	&	0.0230	&	0.0680	\\
	&	-	&	(0.00015)	&	(0.00002)	&	(0.00003)	\\[-6pt]
\multirow{2}{*}{48}	&	0.5430	&	0.5400	&	0.0470	&	0.1160	\\
	&	-	&	(0.00011)	&	(0.00001)	&	(0.00002)	\\[-6pt]
\multirow{2}{*}{80}	&	0.5510	&	0.5480	&	0.0780	&	0.1680	\\
	&	-	&	(0.00008)	&	(0.00001)	&	(0.00002)	\\[-6pt]
		\midrule
\multirow{2}{*}{$\hat{\beta}$}	&	0.1049	&	0.0719 	&	1.0000	&	0.7694	\\
	&	$R^2$=0.8891	&	$R^2$=0.9253	&	$R^2$=1.0000	&	$R^2$=0.9982	\\[-6pt]
			\toprule
	&	\multicolumn{4}{c}{\textbf{Average Path Length} ($\ell$)} \\[-6pt]
\textbf{Avg Deg ($\bar{k}$)}	&	Lattice	&	Small World	&	Erd\"{o}s-Renyi	&	Scale-Free	\\
	\midrule
\multirow{2}{*}{8}	&	10.6820	&	5.0780	&	3.5640	&	3.1780	\\
	&	-	&	(0.00202)	&	(0.00074)	&	(0.00069)	\\[-6pt]
\multirow{2}{*}{24}	&	5.5910	&	3.0100	&	2.5330	&	2.4550	\\
	&	-	&	(0.00036)	&	(0.00019)	&	(0.00016)	\\[-6pt]
\multirow{2}{*}{48}	&	3.8890	&	2.5320	&	2.0530	&	2.0780	\\
	&	-	&	(0.00016)	&	(0.0001)	&	(0.00007)	\\[-6pt]
\multirow{2}{*}{80}	&	3.0300	&	2.1550	&	1.9240	&	1.9350	\\
	&	-	&	(0.00014)	&	(0.00001)	&	(0.00001)	\\	[-6pt]\midrule
\multirow{2}{*}{$\hat{\beta}$}	&	-0.5634	&	-0.3927	&	-0.2860	&	-0.2242	\\
	&	$R^2$=0.9999	&	$R^2$=0.9837 	&	$R^2$=0.9903	&	$R^2$=0.9954	\\[-6pt]

	\bottomrule
	\end{tabular}
	}
		\caption{\label{tab:theor_graph_properties}Expected values of the standard deviation of degree distribution ($s_k$), global clustering coefficient ($c$) and average path length ($\ell$) in the four families of networks under study for average degree ($\bar{k}$) in the range \{8,24,48,80\}. Network size $N=1024$. Avg Deg ($\bar{k}$): Exact average degree for 2-dim lattices with Moore neighborhoods ($LA$) and expected average degree for Small-World ($SW$), Erd\"{o}s-Renyi ($ER$) and Scale-Free ($SF$) networks. Standard errors for Monte Carlo averages with sample size $M=1000$ are reported in parentheses. In the average degree range considered, all four metrics scale with $\bar{k}$ approximately as $\bar{k}^{\beta}$. The $R^2$ of the fit is reported below its maximum-likelihood estimate $\hat{\beta}$. Note also that, for given $\bar{k}\in\{8,24,48,80\}$ the following inequalities hold: $s_k(LA)<s_k(SW)<s_k(ER)<s_k(SF)$, $c(ER)<c(SF)<c(SW)<c(LA)$ and $\ell(SF)<\ell(ER)<\ell(SW)<\ell(LA)$.}

\end{table}

\newpage

%\vfill
\begin{table}[t!]							
	\begin{tabular}{P{3cm}|P{3cm}P{3cm}P{3cm}}						
	\toprule						
	&	\multicolumn{3}{c}{\textbf{Global Parameters}}					\\
	\midrule
 $N$	&	\multicolumn{3}{c}{1024}					\\
$\theta$	&	\multicolumn{3}{c}{0.05}					\\
$D$	&	\multicolumn{3}{c}{Log Normal with $(\mu,\sigma)=(1.621,0.418)$}					\\
	\toprule
	&	\multicolumn{3}{c}{\textbf{Network Parameters}}					\\
	\midrule

$\bar{k}$	&	8	&	24	&	48	\\
$r^{LA}$	&	1	&	2	&	3	\\
$r^{SW}$	&	4	&	12	&	24	\\
$p^{ER}$	&	$8\cdot 1023^{-1}$	&	$24\cdot 1023^{-1}$	&	$48\cdot 1023^{-1}$	\\
$m^{SF}$	&	4	&	12	&	24	\\
	\toprule
	&	\multicolumn{3}{c}{\textbf{Epidemic Scenarios}}					\\
		&	\textit{Strong impact}	&	\textit{Mid Impact}	&	\textit{Low Impact}	\\
		\midrule

$DQR$	&	$0.20$	&	$0.15$	&	$0.10$	\\
$DDR$	&	$0.10$	&	$0.07$	&	$0.04$	\\
$DRR$	&	$0.10$	&	$0.15$	&	$0.20$	\\
$\alpha$	&	$0.20$	&	$0.10$	&	$0.05$	\\
\bottomrule							
\end{tabular}	
\caption{\label{tab:pars_setup}Parametrizations employed in Monte Carlo simulations.} 					
\end{table}							

%\vskip 3cm

\begin{figure}[b!]
	\includegraphics[width=12cm]{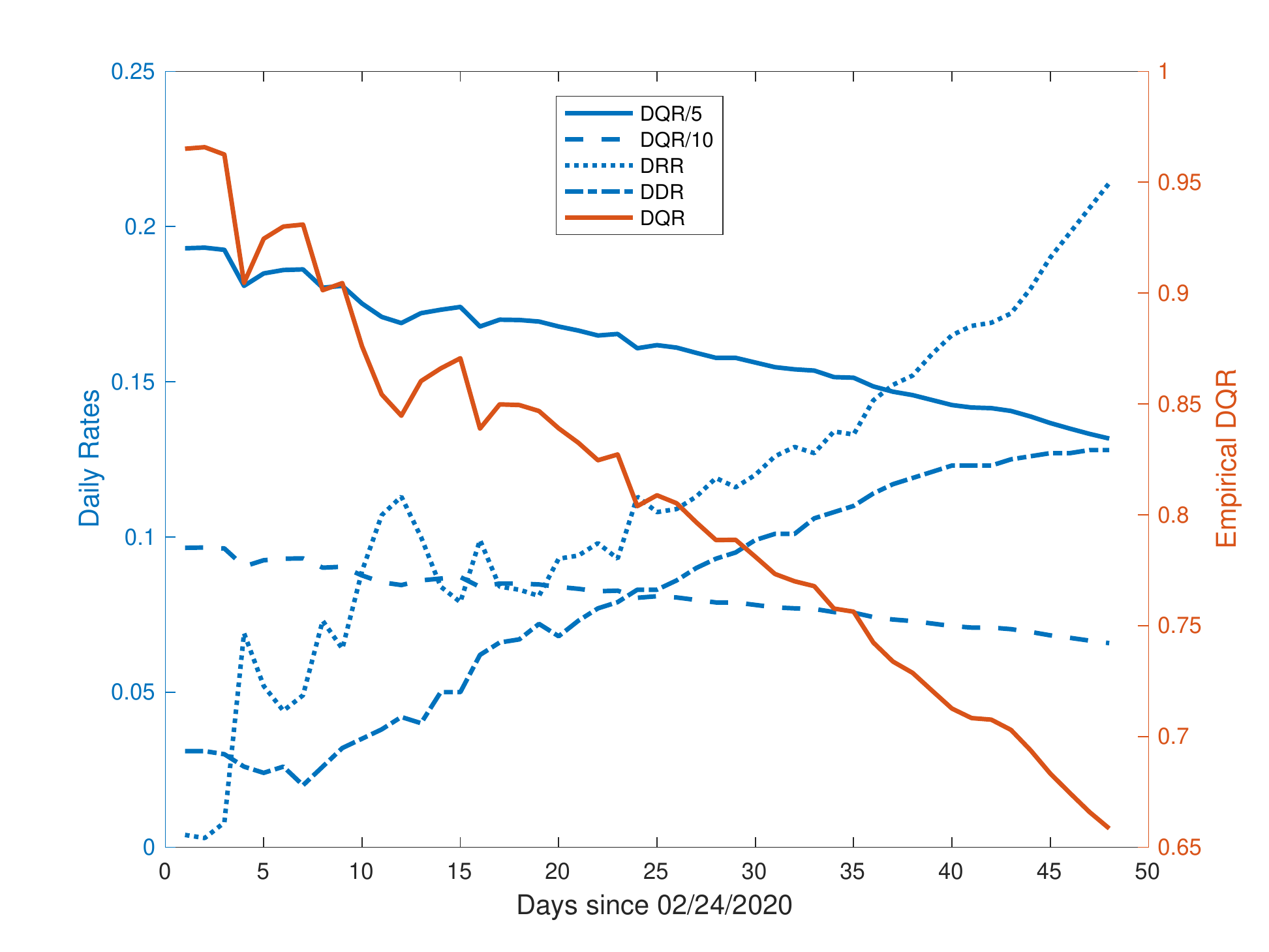}
	\caption{\label{fig:italian_epidemics}Time series of daily quarantined rate, recovered rate and death rate for Italy in the first 48 days of the epidemic diffusion. Data from ``Dipartimento della Protezione Civile'', covering the period from February, 22nd onward. Blue lines: scale on left axis. Red line: scale on right axis. Empirical $DQR$ obtained by dividing quarantined people by total detected cases. $DQR/5$ and $DQR/10$ assume that true infected people are respectively 5 and 10 times higher.}
\end{figure}

\vfill

\begin{figure}[h]
	\centering
	\includegraphics[width=16cm]{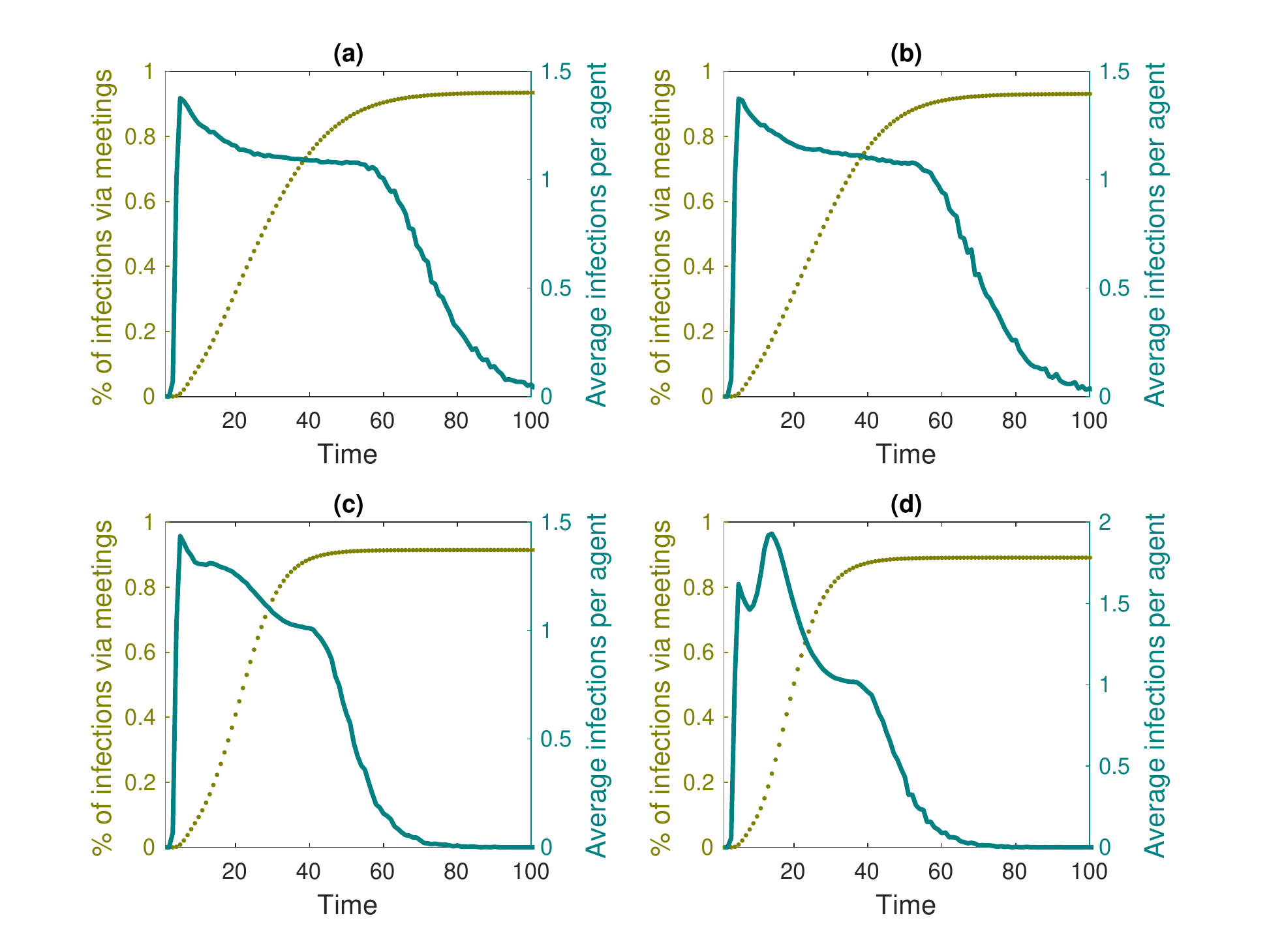}
	\caption{\label{fig:infections}Within-simulation evolution of spatial correlation coefficients of compartments (SCCC). Colors: $S$ (blue), $E$ (yellow), $I$ (red), $Q$ (magenta), $R$ (green) and $D$ (black) over time. Averages across $M$=1000 Monte Carlo simulations. Panels: (a) regular 2-dimensional lattice with Moore neighborhoods; (b) small-world lattice; (c) Erd\"{o}s-Renyi random graph; (d) scale-free network.}
\end{figure}

\newpage

\begin{figure}[h!]
	\centering
	\includegraphics[width=14cm,height=14cm]{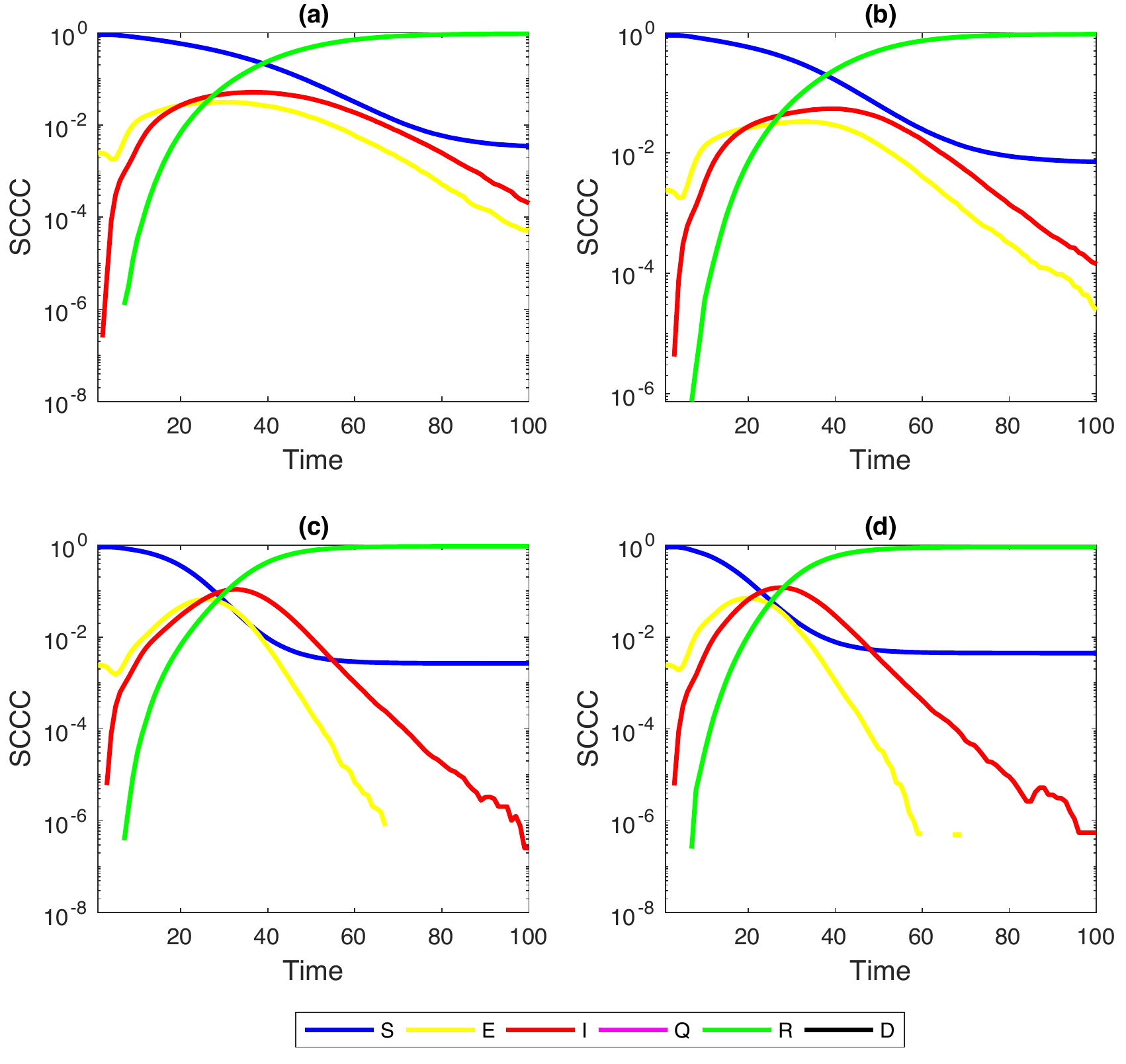}
	\caption{\label{fig:seiqrd_spat_corr}Within-simulation evolution of spatial correlation coefficients of compartments (SCCC). Colors: $S$ (blue), $E$ (yellow), $I$ (red), $Q$ (magenta), $R$ (green) and $D$ (black) over time. Averages across $M$=1000 Monte Carlo simulations. Panels: (a) regular 2-dimensional lattice with Moore neighborhoods; (b) small-world lattice; (c) Erd\"{o}s-Renyi random graph; (d) scale-free network.}
\end{figure}

\newpage

\begin{figure}[ht!]
	\centering
	\includegraphics[width=12cm,keepaspectratio]{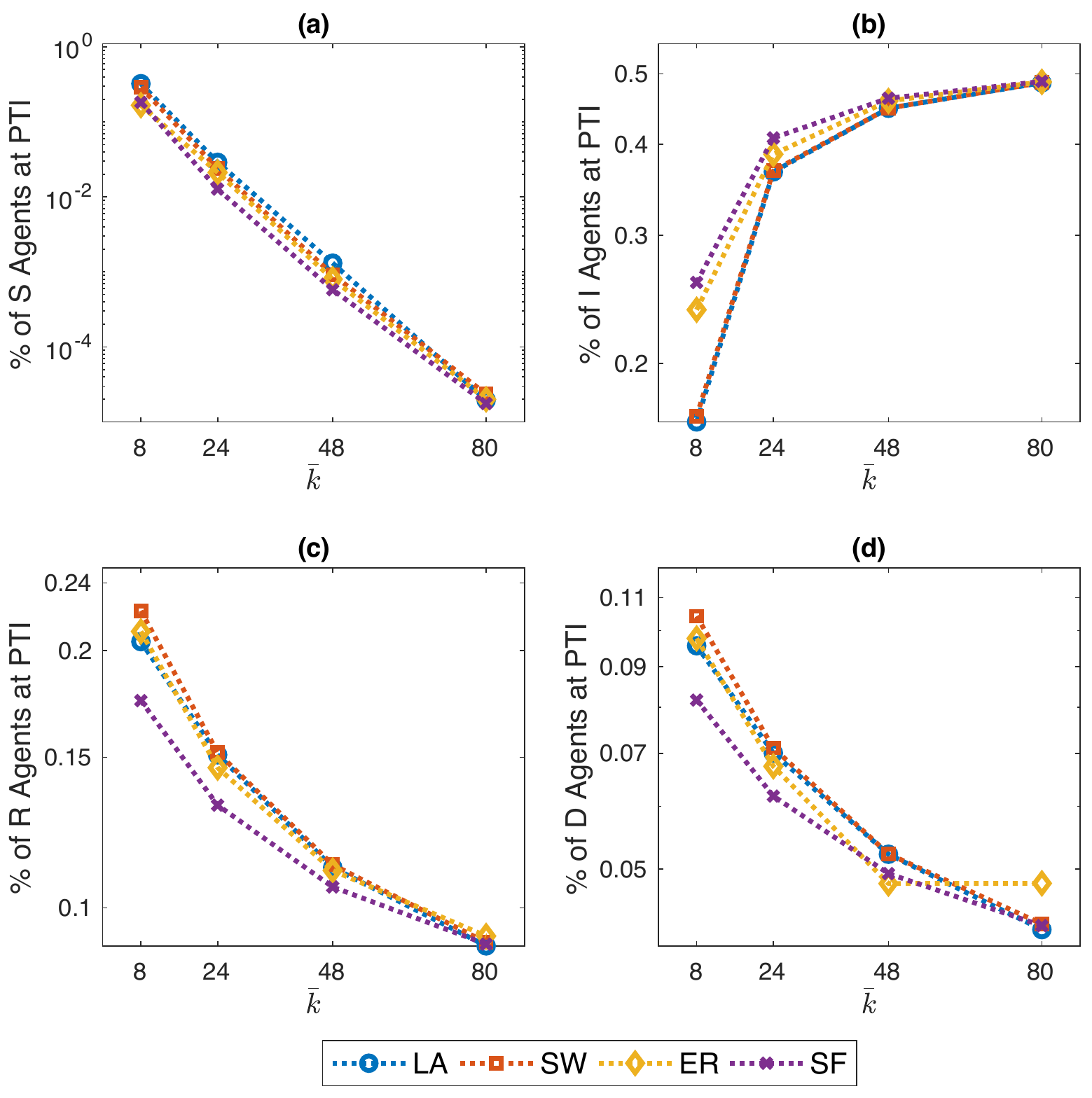}
	\caption{\label{fig:deg_shares}Shares of agents in compartments $S$, $I$, $R$ and $D$ at peak-time of infections (PTI), when initial average degree ranges in $\{8,24,40,80\}$. Mid-impact epidemic scenario. Averages across $M$=1000 Monte Carlo simulations. Y-axis in log scale. }
\end{figure}

\begin{figure}[hb!]
	\centering
	\includegraphics[width=17cm,keepaspectratio]{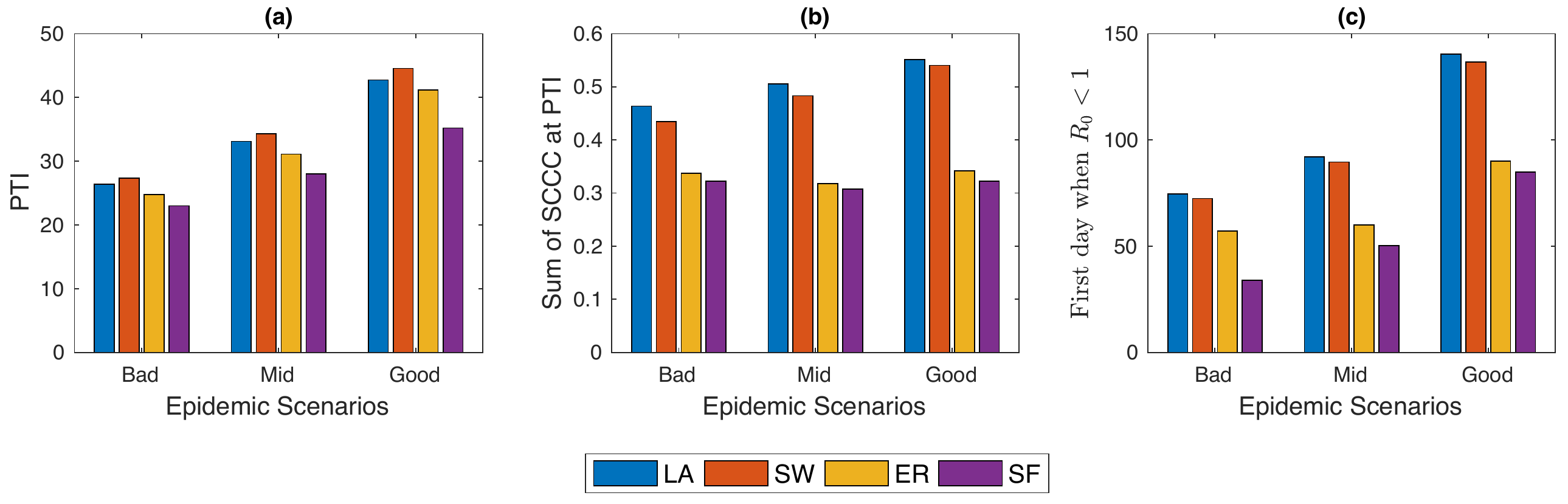}
	\caption{\label{fig:epid_scen_fig2}Comparing the behavior of the model across three epidemic scenarios: Bad vs. Mid vs. Good (see Section \ref{subsec:pars_setup}). Panel (a): Peak-time of infections (PTI). Panel (b): Sum of spatial correlation coefficients of compartments (SCCC) at PTI. Panel (c): estimate of the first day after which $\bar{\rho}$ goes below one (cf. SM, Section \ref{sec:r0}). Initial average degree: $\bar{k}=8$. Averages across $M$=1000 Monte Carlo simulations. }
\end{figure}

\newpage \clearpage
\setstretch{1.2}

\section{Calibration of Epidemic Parameters\label{sec:calibration_epidemics}}

In order to calibrate epidemic parameters for the Italian case, I use data from ``Dipartimento della Protezione Civile'', available at \url{https://github.com/pcm-dpc/COVID-19/tree/master/dati-andamento-nazionale}, covering the period from February, 22nd onward. The time series of total \textit{detected} cases ($TC$) is broken into: (i) currently positive total \textit{detected} cases, i.e. individuals showing severe symptoms that are therefore quarantined ($Q$), either at home or in the hospital; (ii) recovered patients ($R$); (iii) dead patients ($D$). 

Daily quarantined, recovered, and death rates are computed as ratios to total \textit{detected} cases ($TC$), see Figure \ref{fig:italian_epidemics}.  Since this figure is highly dependent on the number of swabs taken and asymptomatic individuals, the daily quarantined rate ($DQR$), in particular, is likely to suffer from strong overestimation. Indeed, several contributions have shown that for Italy actual infected people may be 5-10 times higher \cite{Fenga_2020}. Therefore, in the figure we also report the $DQR$ for these overestimation scenarios. We do not rescale $DRR$ and $DDR$ accordingly as we are here interested in the fraction of people who, after being quarantined, transition in the $R$ and $Q$ compartments.

Simulation scenarios in Table \ref{tab:pars_setup} are then built using the range of $DQR_5$, $DQR_10$,$DRR$ and $DDR$.            

\section{Estimating the Basic Reproduction Number ($R_0$)\label{sec:r0}}

In each day, agent $i$ in state $S$ may possibly become $E$ if h/she meets $k\geq 1$ neighbors $j\in\{j_1,\dots,j_k\}\subset P$ in state $I$. If this happens, the burden of having infected $i$ is divided in equal shares $\eta_{j_h}=1/k$ across all infecting neighbors. If an $I$ agent contributes over the day to the infection of more than one neighbor, all its $\eta$ shares are accordingly cumulated. At the end of each day, after all infections have been eventually occurred, we compute $\bar{\rho}$ as the population average (over non-zero elements) of $\eta$ shares.   

The $\bar{\rho}$ statistics can be interpreted as a rough estimate of the ``basic reproduction number'' ($R_0$) of the epidemic process, as it measures the average number of cases that each infected agent generates.

Given a single simulation, the time-series $\bar{\rho}_t$ typically goes down with $t$, as the set of infective agents shrinks and the number of recovered (or dead) people increases. It is therefore worthwhile to estimate $t^\ast = \min\{1 \leq t \leq T: \bar{\rho}_t <1\}$, which is computed at the end of each simulation as one of its summarizing statistics. Since in each simulation $\bar{\rho}_t$ may oscillate many times around one before decreasing persistently below one, I fit the series with four different functions: (i) polynomial of degree 1: $f(t)=at+b$; (ii) polynomial of degree 2: $f(t)=at^2+bt+c$; (ii) power: $f(t)=at^b$; (iii) exponential $f(t)=a\exp{(-bt)}$; and then for the best fit $f^\ast$ (according to the associated adjusted $R^2$), I define $t^\ast=\min\{1 \leq t \leq T: f^\ast(t) <1\}$ for $t=1,\dots,T$.  	

\section{Network Metrics vs. $Q$ and $R$ Shares\label{sec:QD_NetStats}}
\begin{figure}[b!]
	\centering
	\includegraphics[width=16cm,keepaspectratio]{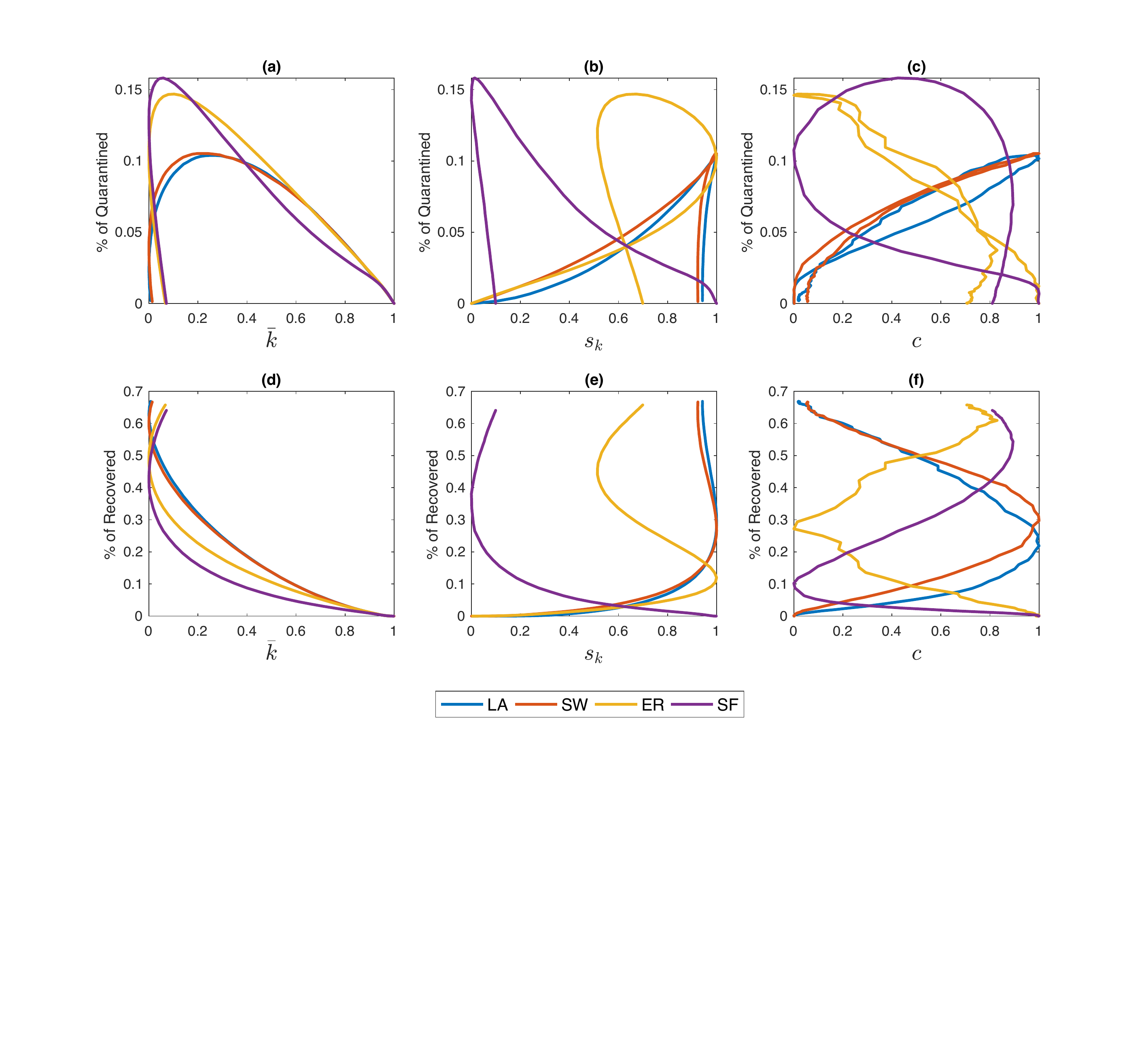}
	\caption{\label{fig:QD_NetStats}Within-simulation evolution of network metrics, re-scaled to match the $[0,1]$ interval, against the shares of $Q$ and $R$ agents. Initial $\bar{k}=8$. Mid-impact epidemic scenario. Panels (a)-(c): Population share of $Q$ agents. Panels (d)-(f): Population share of $R$ agents. Averages across $M$=1000 Monte Carlo simulations.}
\end{figure}

This Section reports more evidence about how the evolution of $Q$ and $R$ shares co-evolve, within a simulation, with the structure of the network where agents are embedded in. I still focus on a `Mid Impact' epidemic scenario, with $\bar{k}=8$ for convenience, and plot in Figure \ref{fig:QD_NetStats} the within-simulation time series of network metrics, re-scaled to match the unit interval, against the shares of $Q$ and $R$ agents. Average path length ($\ell$) is not taken into consideration here as its relationship with $\bar{k}$ is monotone along the process.

A quick inspection of Figure \ref{fig:QD_NetStats} suggests the process undergoes a series of phases that depend on the family of network considered. These phases are summarized in the following tables, where for each relationship between network metrics $(\bar{k},s_k,c)$, shares of agents in $Q$ and $R$ states, and network structure, I identify phases (i.e., subsets of time-series evolution) where they display a particular co-movement. For example, a pair $(\uparrow,\downarrow)$ for the entry $(x,y)$ means a phase where the network metrics $x$ increases and the share of agent in state $y$ decreases.   

\vskip 1cm

\begin{table*}[ht!]

	\begin{tabular}{c|c|c|c|c|}
		$(\bar{k},\% $Q$)$ & $LA$ & $SW$ & $ER$ & $SF$ \\
		
		\toprule
		Phase I & \multicolumn{4}{|c|}{($\downarrow,\uparrow$)} \\ \midrule
		Phase II & \multicolumn{4}{|c|}{($\downarrow,\downarrow$)} \\ \midrule
		Phase III & \multicolumn{2}{|c|}{} & \multicolumn{2}{|c|}{($\uparrow,\downarrow$)}\\
		\bottomrule		
	\end{tabular}
\quad \quad \quad 
	\begin{tabular}{c|c|c|c|c|}
		$(\bar{k},\% $R$)$ & $LA$ & $SW$ & $ER$ & $SF$ \\
		
		\toprule
		Phase I & \multicolumn{4}{|c|}{($\downarrow,\uparrow$)} \\ \midrule
		Phase II & \multicolumn{4}{|c|}{($\uparrow,\uparrow$)} \\ \bottomrule
%		 & \multicolumn{2}{c}{} & \multicolumn{2}{c}{}\\
	\end{tabular}

\end{table*}
	
\begin{table*}[ht!]

	\begin{tabular}{c|c|c|c|c|}
		$(s_{k},\% $Q$)$ & $LA$ & $SW$ & $ER$ & $SF$ \\
		
		\toprule
		Phase I & \multicolumn{3}{|c|}{($\uparrow,\uparrow$)} & ($\downarrow,\uparrow$) \\ \midrule
		Phase II & \multicolumn{2}{|c|}{($\downarrow,\downarrow$)} & ($\downarrow,\uparrow$) & ($\uparrow,\downarrow$)\\ \midrule
		Phase III & \multicolumn{2}{|c|}{} & ($\uparrow,\uparrow$) & \\ \midrule
		Phase IV & \multicolumn{2}{|c|}{} & ($\uparrow,\downarrow$) & \\
		\bottomrule		
	\end{tabular}
\quad \quad
	\begin{tabular}{c|c|c|c|c|}
		$(s_{k},\% $R$)$ & $LA$ & $SW$ & $ER$ & $SF$ \\
		
		\toprule
		Phase I & \multicolumn{3}{|c|}{($\uparrow,\uparrow$)} & ($\downarrow,\uparrow$) \\ \midrule
		Phase II & \multicolumn{3}{|c|}{($\downarrow,\uparrow$)} & ($\uparrow,\uparrow$) \\ \midrule
		Phase III & \multicolumn{2}{|c|}{} & ($\uparrow,\uparrow$) & \\
		\bottomrule		
	\end{tabular}

\end{table*}

\begin{table}[ht!]

	\begin{tabular}{c|c|c|c|c|}
		$(s_{k},\% $Q$)$ & $LA$ & $SW$ & $ER$ & $SF$ \\
		
		\toprule
		Phase I & \multicolumn{2}{|c|}{($\uparrow,\uparrow$)} & \multicolumn{2}{|c|}{($\downarrow,\uparrow$)} \\ \midrule
		Phase II & \multicolumn{2}{|c|}{($\downarrow,\downarrow$)} & ($\uparrow,\downarrow$) & ($\uparrow,\uparrow$)\\ \midrule
		Phase III & \multicolumn{2}{|c|}{} & ($\downarrow,\downarrow$) & ($\uparrow,\downarrow$) \\ \midrule
		Phase IV & \multicolumn{2}{|c|}{} &  & ($\downarrow,\downarrow$)\\ 
		\bottomrule		
	\end{tabular}
\quad \quad
	\begin{tabular}{c|c|c|c|c|}
		$(s_{k},\% $R$)$ & $LA$ & $SW$ & $ER$ & $SF$ \\
		
		\toprule
		Phase I & \multicolumn{2}{|c|}{($\uparrow,\uparrow$)} & \multicolumn{2}{|c|}{($\downarrow,\uparrow$)} \\ \midrule
		Phase II & \multicolumn{2}{|c|}{($\downarrow,\uparrow$)} & \multicolumn{2}{|c|}{($\uparrow,\uparrow$)} \\ \midrule
		Phase III & \multicolumn{2}{|c|}{} &  \multicolumn{2}{|c|}{($\downarrow,\uparrow$)} \\
		\bottomrule		
	\end{tabular}

\end{table}

\vskip 1cm

\end{appendices}

\end{document}